\documentclass{radioeng}

\usepackage[czech,english]{babel}
\usepackage[cp1250]{inputenc}
\usepackage{graphics}

\usepackage{graphicx}
\usepackage{epsfig}
\usepackage{dcolumn}
\usepackage{bm}
\usepackage{amsmath}
\usepackage{amssymb}

\usepackage{tabularx}
\usepackage{textcomp}

\usepackage{multicol}

\usepackage{listings}

\newcommand\Small{\fontsize{8.3}{8.3}\selectfont}

\begin{document}

%%%%%%%%%%%%%%%%%%%%%%%%%%%%%%%%%%%%%%%%%%%%%%%%%%%%%%%%%%%%%%%
% Header, title, authors, affiliation
%%%%%%%%%%%%%%%%%%%%%%%%%%%%%%%%%%%%%%%%%%%%%%%%%%%%%%%%%%%%%%%

\setcounter{page}{11}

\rengHeader{XX}{YY}{April 2013}{Y. V. PERSHIN, M. DI VENTRA, SPICE MODEL OF MEMRISTIVE DEVICES WITH THRESHOLD}

\rengTitle{SPICE model of memristive devices with threshold}

\rengNames{Yuriy V. PERSHIN$^1$, Massimiliano DI VENTRA$^2$}

\rengAffil{$^1$Department of Physics and Astronomy and USC Nanocenter, University of South Carolina, Columbia, SC 29208, USA \\
$^2$Department of Physics, University of California, San Diego, La Jolla, CA 92093-0319, USA}

\rengMail{pershin@physics.sc.edu, diventra@physics.ucsd.edu}

%%%%%%%%%%%%%%%%%%%%%%%%%%%%%%%%%%%%%%%%%%%%%%%%%%%%%%%%%%%%%%%
% Abstract, keywords
%%%%%%%%%%%%%%%%%%%%%%%%%%%%%%%%%%%%%%%%%%%%%%%%%%%%%%%%%%%%%%%
\begin{multicols}{2}

\begin{rengAbstract}
Although memristive devices with threshold voltages are the norm rather than the
exception in experimentally realizable systems, their SPICE programming is not yet common.
Here, we show how to implement such systems in the SPICE environment. Specifically, we present SPICE models
of a popular voltage-controlled memristive system specified by five different parameters for PSPICE and NGSPICE circuit simulators. We expect this
implementation to find widespread use in circuits design and testing.
\end{rengAbstract}

\rengKeywords{Memristive devices, memristor, memristor model, threshold dynamics}

%%%%%%%%%%%%%%%%%%%%%%%%%%%%%%%%%%%%%%%%%%%%%%%%%%%%%%%%%%%%%%%
% Paper text
%%%%%%%%%%%%%%%%%%%%%%%%%%%%%%%%%%%%%%%%%%%%%%%%%%%%%%%%%%%%%%%

\rengSection{Introduction}

In the last few years, circuit elements with memory, namely, memristive \cite{chua76a}, memcapacitive and meminductive \cite{diventra09a} systems have attracted considerable attention from different disciplines due to their capability of non-volatile low-power information storage, potential applications in analog and digital circuits, and their ability to store and manipulate information on the same physical platform \cite{pershin11a}. However, when combined into complex circuits, progress in this field significantly relies on the available tools at our disposal. One such tool is the SPICE simulation environment, commonly used in circuit simulations and testing.  While several SPICE models of memristive \cite{Biolek2009-1,Benderli2009-1,Biolek2009-2,Shin10a,Rak10a,Yakopcic11a,Kvatinsky12a}, memcapacitive \cite{Biolek2009-2,Biolek10b} and meminductive \cite{Biolek2009-2,Biolek11b} elements are already available, they typically \cite{Biolek2009-1,Benderli2009-1,Biolek2009-2,Shin10a,Rak10a} rely on physical models without a threshold (see, e.g., Refs. \cite{strukov08a,joglekar09a}).

Threshold-type switching is instead an extremely important common feature of memristive devices (for examples, see Ref. \cite{pershin11a}) and, due to physical constraints, likely to be common in
memcapacitive and meminductive elements as well \cite{diventra13a}. Indeed, it is the threshold-type switching which is responsible for non-volatile information storage, serves as a basis for logic operations \cite{borghetti10a,pershin12a}, etc., and therefore, it can not be neglected. For instance, experimentally demonstrated memristive logic circuits \cite{borghetti10a} and emerging memory architectures \cite{linn10a} support fixed-threshold modeling \cite{pershin09b} of memristive devices. Moreover, the atomic migration responsible for resistance switching in many important experimental systems is induced by the applied field and not by the electric current flow. Therefore, models with voltage threshold \cite{pershin09b,Yakopcic11a} are physically better justified than those with the current one \cite{Kvatinsky12a}.

In the present paper we introduce a SPICE model for a memristive device with threshold voltage that has been proposed by the present authors \cite{pershin09b}. Using this type of memristive devices, we have already demonstrated and analyzed several electronic circuits including a learning circuit \cite{pershin09b}, memristive neural networks \cite{pershin10c}, logic circuits \cite{pershin12a}, analog circuits \cite{pershin10d} and circuits transforming memristive response into memcapacitive and meminductive ones \cite{pershin09e}. These previous results thus demonstrate the
range of applicability of the selected physical model. As a consequence, we expect its SPICE implementation to find numerous applications as well.

\begin{figure}
 \begin{center}
 \includegraphics[angle=0,width=6.5cm]{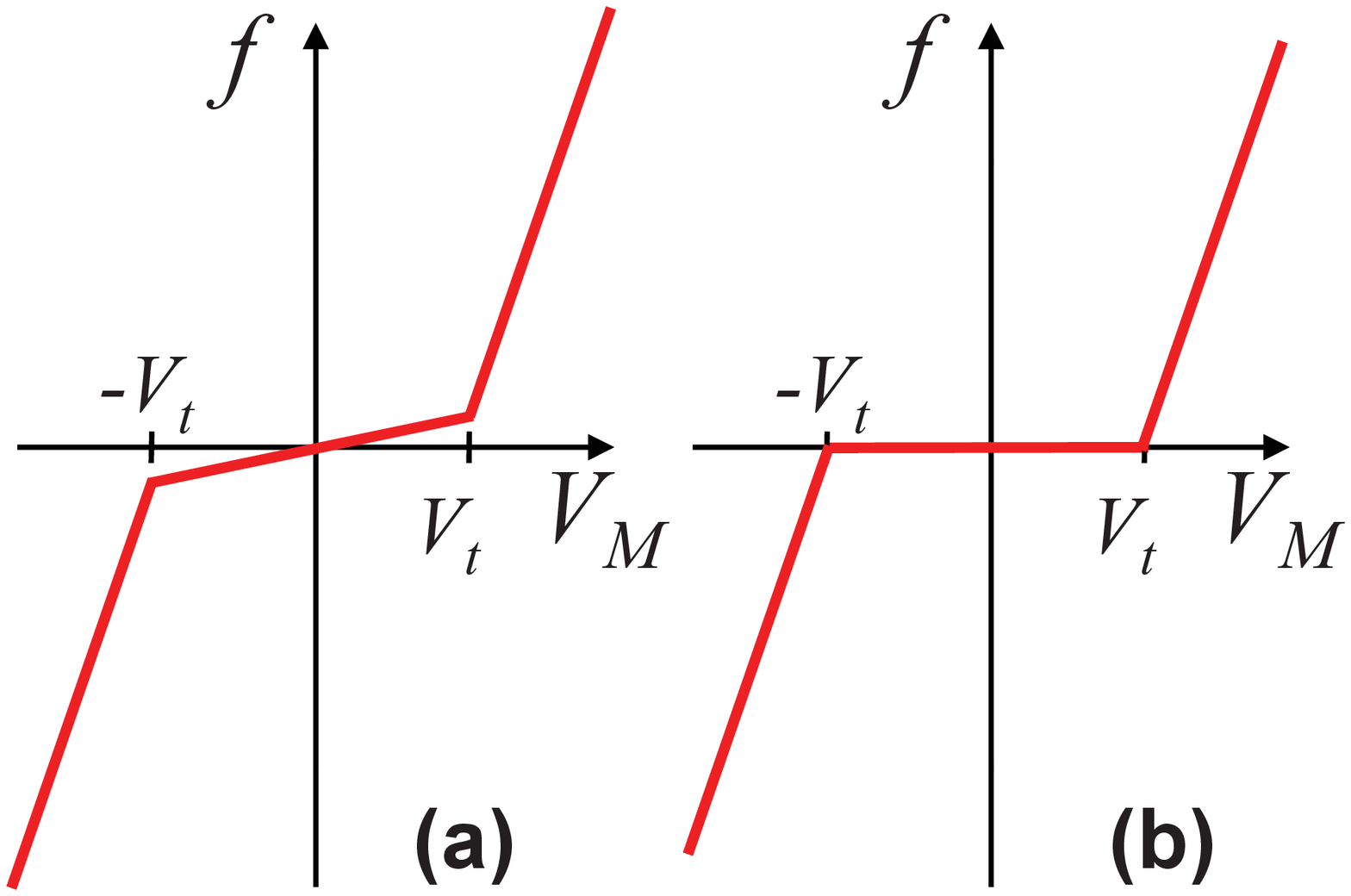}
\fcaption{\label{fig1} Sketch of the function $f(V_M)$ for $\textnormal{(a)}$ $\alpha>0$ and $\beta>0$ and $\textnormal{(b)}$ $\alpha=0$ and $\beta>0$.}
 \end{center}
\end{figure}

\rengSection{SPICE model}

The equations describing memristive systems can be formulated in the voltage- or current-controlled form \cite{chua76a}.
In some cases, a voltage-controlled memristive system can be easily re-formulated as a current-controlled one and vice versa \cite{pershin11a}. Let us then focus on voltage-controlled memristive systems whose general definition (for an $n$th-order voltage-controlled memristive system) is given by the following relations
\begin{eqnarray}
I(t)&=&R_M^{-1}\left(X,V_M,t \right)V_M(t) , \label{Condeq1}\\
\dot{X}&=&f\left( X,V_M,t\right) \label{Condeq2}
\end{eqnarray}
where $X$ is the vector representing $n$ internal state variables,
$V_M(t)$ and $I(t)$ denote the voltage and current across the
device, and $R_M$ is a scalar, called the {\em memristance} (for
memory resistance).

\begin{figure} [t]
 \begin{center}
 \includegraphics[angle=0,width=5.5cm]{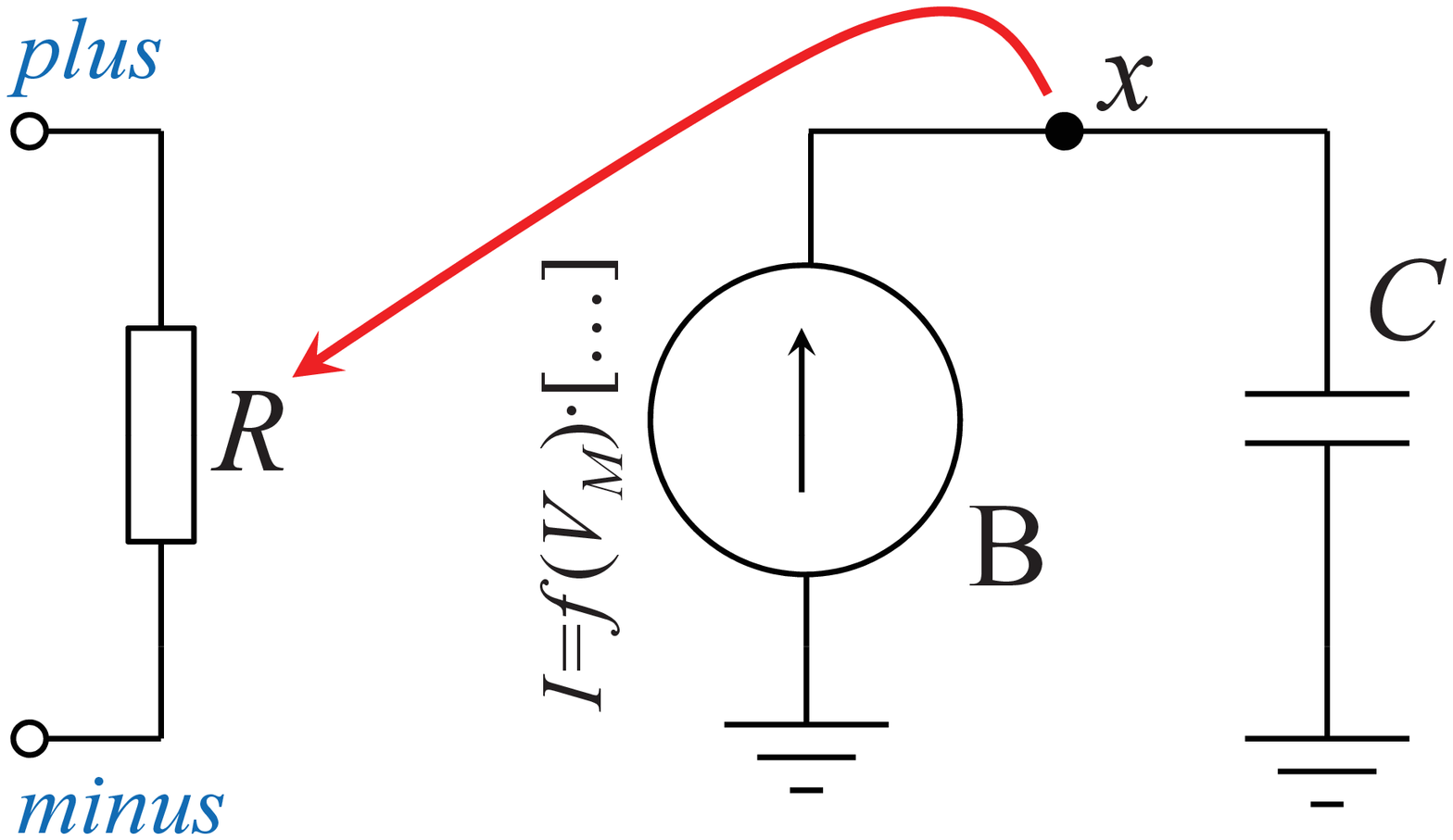}
\fcaption{\label{fig2}  Schematic of the SPICE model. The memristive device functionality is organized as a subcircuit consisting of a behavioral resistor $R$,  current source B and capacitor $C$. The voltage across the capacitor (at the node $x$) defines the resistance of $R$.}
 \end{center}
\end{figure}

\begin{table*}
\begin{center}
\begin{lstlisting}
.subckt memristor pl mn PARAMS: Ron=1K Roff=10K Rinit=5K alpha=0 beta=1E13 Vt=4.6
Bx 0 x I='(f1(V(pl,mn))>0) && (V(x)<Roff) ? {f1(V(pl,mn))}: (f1(V(pl,mn))<0) && (V(x)>Ron) ? {f1(V(pl,mn))}: {0}'
Cx x 0 1 IC={Rinit}
R0 pl mn 1E12
Rmem pl mn r={V(x)}
.func f1(y)={beta*y+0.5*(alpha-beta)*(abs(y+Vt)-abs(y-Vt))}
.ends
\end{lstlisting}
\end{center}
\tcaption{\label{tbl1} NGSPICE implementation of Eqs. (\ref{eq3})-(\ref{eq5}).}
\end{table*}

\begin{table*}
\begin{center}
\begin{lstlisting}
.subckt memristor pl mn PARAMS: Ron=1K Roff=10K Rinit=5K beta=1E13 Vtp=4.6 Vtm=4.6 nu1=0.0001 nu2=0.1
Gx 0 x value={f1(V(pl)-V(mn))*(f2(f1(V(pl)-V(mn)))*f3(Roff-V(x))+f2(-f1(V(pl)-V(mn)))*f3(V(x)-Ron))}
Raux x 0 1E12
Cx x 0 1 IC={Rinit}
Gpm pl mn value={(V(pl)-V(mn))/V(x)}
.func f1(y)={beta*(y-Vtp)/(exp(-(y-Vtp)/nu1)+1)+beta*(y+Vtm)/(exp(-(-y-Vtm)/nu1)+1)}
.func f2(y1)={1/(exp(-y1/nu1)+1)}
.func f3(y)={1/(exp(-y/nu2)+1)}
.ends
\end{lstlisting}
\end{center}
\tcaption{\label{tbl2} PSPICE implementation of Eqs. (\ref{eq3})-(\ref{eq5}) for $\alpha=0$. For the sake of versatility,
$Vtp$ and $Vtm$ are introduced to define voltage thresholds separately for opposite polarities.}
\end{table*}

A specific realization of a voltage-controlled memristive system {\it with threshold} has been suggested by the
present authors in Ref. \cite{pershin09b}. Such a memristive system is described by
\begin{eqnarray}
I&=&X^{-1}V_M, \label{eq3} \\
\frac{\textnormal{d}X}{\textnormal{d}t}&=&f\left( V_M\right) \left[
\nonumber \theta\left( V_M\right)\theta\left( R_{off}-X\right) + \right. \\
 &{} &\qquad \qquad \qquad \left.
\theta\left(-V_M\right)\theta\left( X-R_{on}\right)\right], \label{eq4}
\end{eqnarray}
with
\begin{equation}
f(V_M)=\beta V_M+0.5\left( \alpha-\beta\right)\left[ |V_M+V_t|-|V_M-V_t| \right] \label{eq5}
\end{equation}
\vskip 1mm
\noindent where $V_t$ is the threshold voltage, $R_{on}$ and $R_{off}$ are limiting values of the memristance $R_M\equiv X$, and the $\theta$-functions (step functions) are used to limit the memristance to the region between $R_{on}$ and $R_{off}$. The important model parameters are the coefficients
$\alpha$  and $\beta$ that characterize the rate of memristance change at $|V_M|< V_t$ and
$|V_M|> V_t$, respectively. These two coefficients define the slopes
of the $f(V_M)$ curve below and above the threshold (see Fig. \ref{fig1}).
When $\alpha=0$ (Fig. \ref{fig1}(b)), the device state changes only if $\left| V_M \right|>V_t$.
Note that Eqs. (\ref{eq3})-(\ref{eq5}) are written in such a way that a positive/negative voltage applied to the top terminal with respect to the bottom
terminal denoted by the black thick line always tends to increase/decrease the memristance $R_M$ (the opposite convention has been used in Ref. \cite{pershin09b}).

\begin{figure}
 \begin{center}
 \includegraphics[angle=0,width=4.5cm]{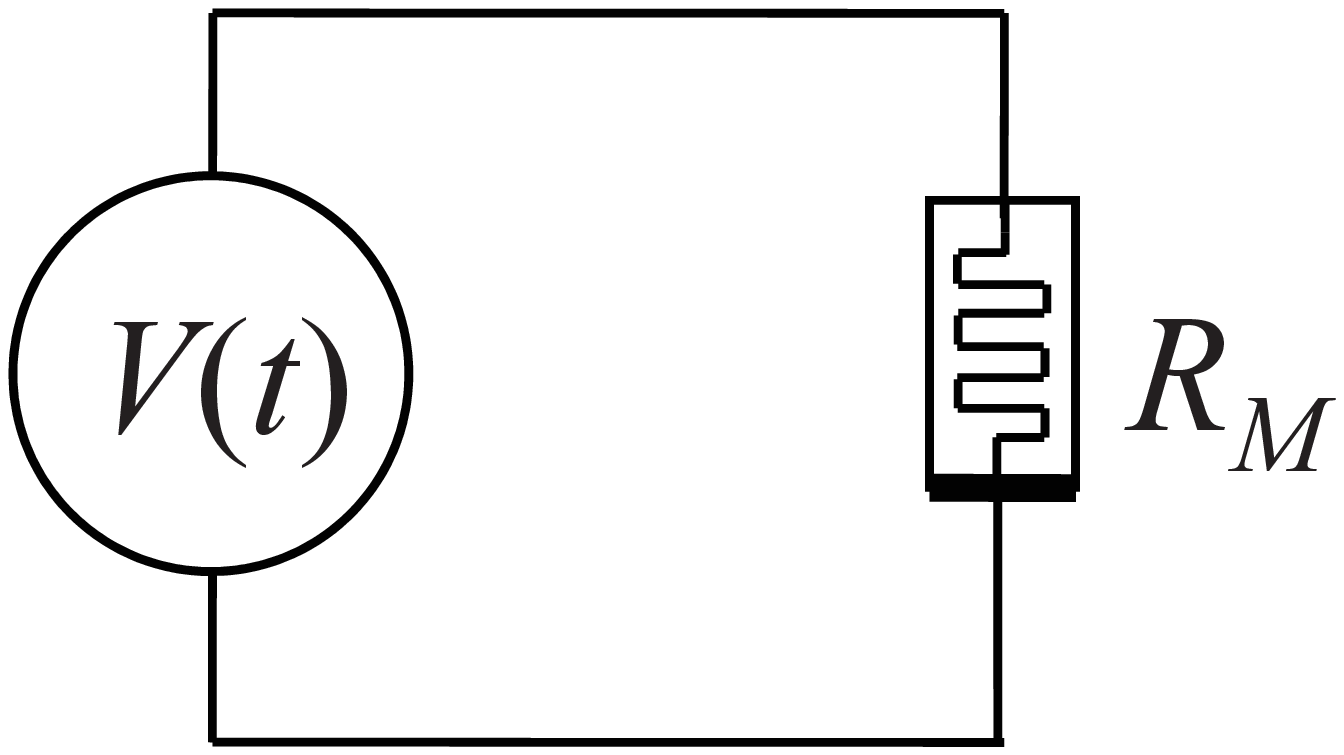}
\fcaption{\label{fig3} Memristive device directly connected to a voltage source $V(t)$.}
 \end{center}
\end{figure}

The SPICE model for these devices is formulated following the general idea of Ref. \cite{Biolek2009-1}. For NGSPICE circuit simulator, the memristive system is realized as a sub-circuit combining a behavioral resistor $R$ (a resistor whose resistance can be specified by an expression), a current source $\uparrow$, and a capacitor $C$.
Table \ref{tbl1} presents the code of the sub-circuit. Its second line ({\it Bx ...}) defines the current source with the current specified through
ternary functions. (A ternary function is defined in the code as $a$ ? $b$ : $c$ , which means "IF a, THEN b, ELSE c" \cite{ngspice}.) The purpose of these functions is to limit $R_M$ between $R_{on}$ and $R_{off}$. The third line of the code in Table \ref{tbl1} specifies the capacitor $C$ ({\it Cx ...}) with
an initial condition. The fourth line ({\it Rmem ...}) defines the behavioral resistor whose resistance takes the same numerical value as the voltage across the capacitor. The next line ({\it .func ...}) provides the function $f$ according to Eq. (\ref{eq5}). We have not experienced any convergence problems using Table \ref{tbl1} model with NGSPICE simulator that potentially could result from "IF-THEN" statements. Clearly, Eq. (\ref{eq3})-(\ref{eq5}) could be programmed differently, employing smoothing functions (e.g., arctan(), sigmoid or similar function) as we do below in the case of PSPICE simulator model. Moreover, instead of "IF...THEN" statement in the Table \ref{tbl1}, one can use a step function based expression.  In this case, the "Bx ..." line of the code should be replaced with "Bx 0 x I=$\{$f1(V(pl,mn))*(u(V(pl,mn))*u(Roff-V(x))+u(V(mn,pl))*u(V(x)-Ron))$\}$".

For PSPICE circuit simulator, the SPICE model of memristive device with threshold is formulated slightly differently
without the use of behavioral resistor. Instead, we employ an additional current source playing the role of behavioral resistor \cite{Basso05a}.
In addition, in order to avoid convergence  problems, the function $f$ in Eq. (\ref{eq5}) should be smoothed. In the most important
case of $\alpha=0$, the smoothing of $f$ is straightforward. Table \ref{tbl2} presents the code for PSPICE circuit simulator for this case.
In Table \ref{tbl2}, $nu1$ and $nu2$ are smoothing parameters used in smoothed step functions $f2$ and $f3$ (although we prefer to use different smoothing parameters for functions of voltages and resistances, a common smoothing function could also be used). We have verified that simulation results are identical in both versions of SPICE and that PSPICE code is also compatible with LTspice circuit simulator. In addition, we note that the value of $beta$ in Table \ref{tbl2} was selected to match switching times of real memristive devices that are in nanoseconds range. We suggest to select the maximum allowable time step not exceeding 0.01ns when using this value of $beta$.

\rengSection{Example}

Let us consider a memristive device with threshold directly connected to a sinusoidal voltage source $V(t)=V_0 \sin(2\pi \nu t)$ as presented in Fig. \ref{fig3}. The circuit simulations are performed as a transient analysis of the circuit taking into account initial conditions (the {\it uic} option of {\it .tran}) within the NGSPICE circuit simulator. In our simulations, we consider two different types of memristive devices with threshold corresponding to two cases of functions $f(V_M)$  as presented in Fig. \ref{fig1}. In the first case (that can be dubbed as a memristive device with a {\it soft threshold}) the coefficients $\alpha,\beta>0$ and $\alpha <\beta$. In this case, the memristance changes at any $V\neq 0$. However, the change is faster when the applied voltage magnitude is above the threshold voltage ($|V|>V_t$). In the second case (Fig. \ref{fig1}(b)), $\alpha =0$. Consequently, the memristance changes only when the applied voltage exceeds the threshold voltage ($|V|>V_{t}$). This second case is closer to the actual behavior of many experimentally realizable memristive systems \cite{pershin11a}. We call this type of systems as memristive devices with {\it hard threshold}.

\begin{figure*}
 \begin{center}
  \centerline{
    \mbox{\includegraphics[width=7.00cm]{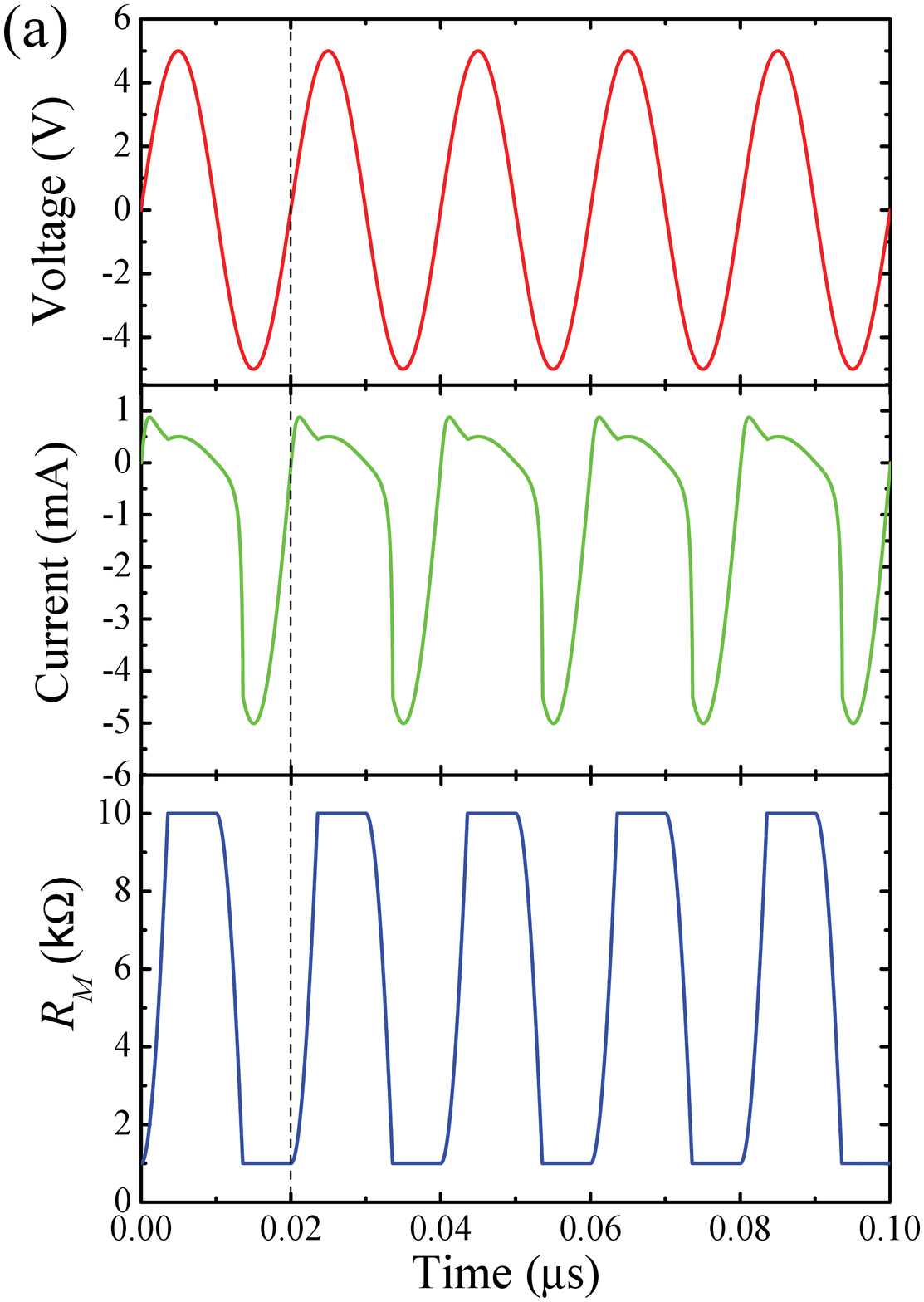}}
    \mbox{\includegraphics[width=7.00cm]{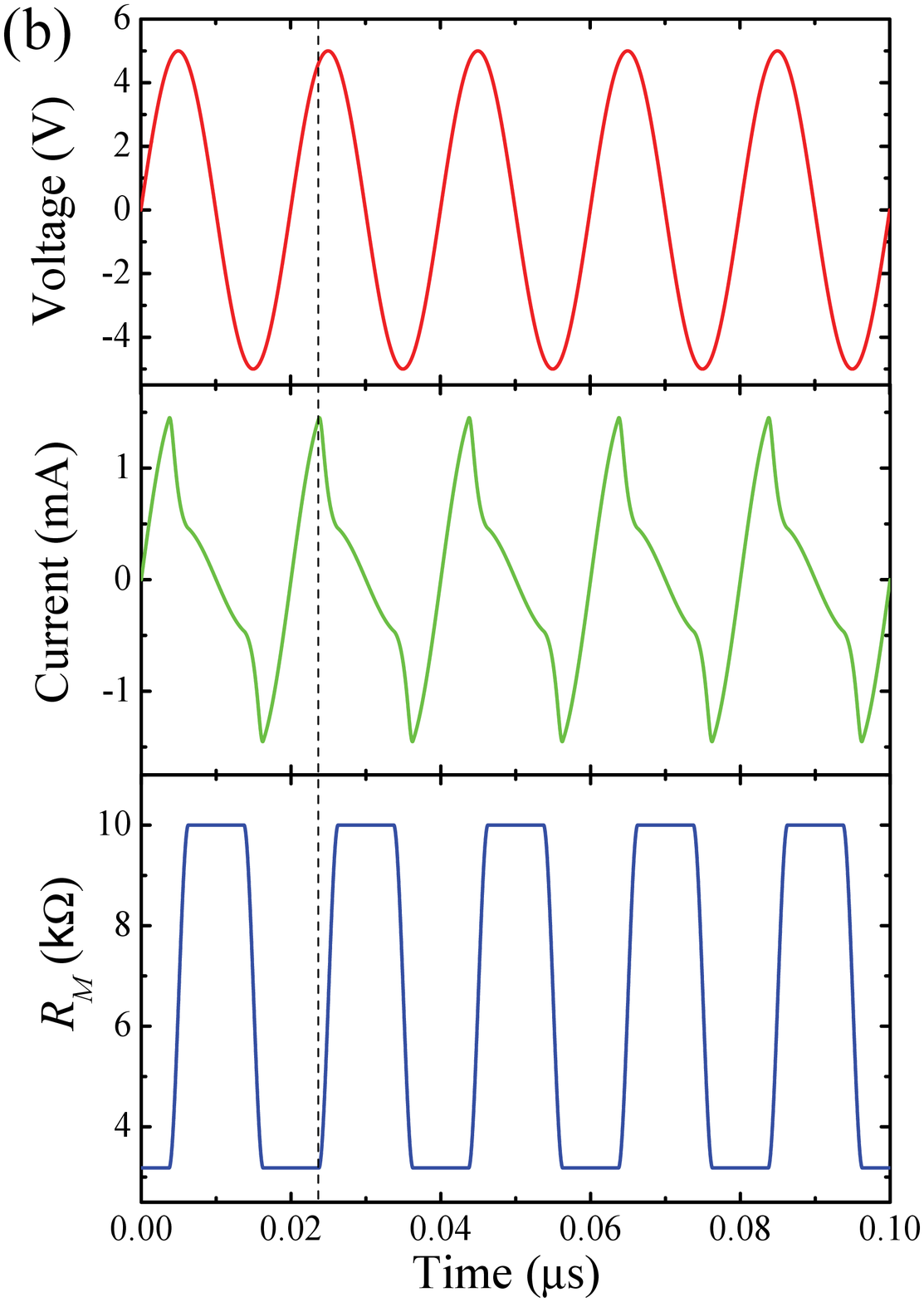}}
   }
\fcaption{ Time evolution of the applied voltage $V(t)$, current $I(t)$ and memristance $R_M(t)$ for memristive devices with (a) soft ($\alpha =0.1\beta$), and (b) hard ($\alpha=0$) thresholds. The vertical dashed lines - corresponding to the onset of switching - serve as guide to the eye. The simulations parameters are as follows: the applied voltage amplitude $V_0=5$V, $\nu=0.05$GHz, $R_{on}=1$k$\Omega$, $R_{off}=10$k$\Omega$, $R_M(t=0)=5$k$\Omega$, $\beta=10^{10}$k$\Omega$/(V s), $V_t=4.6$V.} \label{fig4}
\end{center}
\end{figure*}

Fig. \ref{fig4} presents selected results of our simulations showing the circuit dynamics at long times (the initial transient interval is omitted). We consider two types of memristive devices -- one with a soft and another with hard thresholds ($\alpha=0.1\beta$ and $\alpha=0$, respectively) -- and plot the applied voltage, current and memristance as functions of time for these two cases at a frequency $\nu=0.05$GHz. Clearly, in both cases, the current through the device is not of the simple sine form. The plot of memristance as a function of time demonstrates that the range of memristance change in (a) is larger than in (b) (actually, in (a), $R_M$ switches between $R_{on}$ and $R_{off}$). The vertical dashed line in Fig. \ref{fig4}(a) helps noticing that in Fig. \ref{fig4}(a) the memristance starts changing as soon as the sign of applied voltage changes. In Fig. \ref{fig4}(b), instead, the change of $R_M$ occurs solely when $|V|>V_t$. As a consequence, the shapes of
$R_M(t)$ in Fig. \ref{fig4}(a) and (b) are slightly different, and the steps in
 $R_M(t)$ in Fig. \ref{fig4}(b) are shifted along the horizontal axis compared to those in Fig. \ref{fig4}(a).

\begin{figure*}
 \begin{center}
  \centerline{
    \mbox{\includegraphics[width=7.00cm]{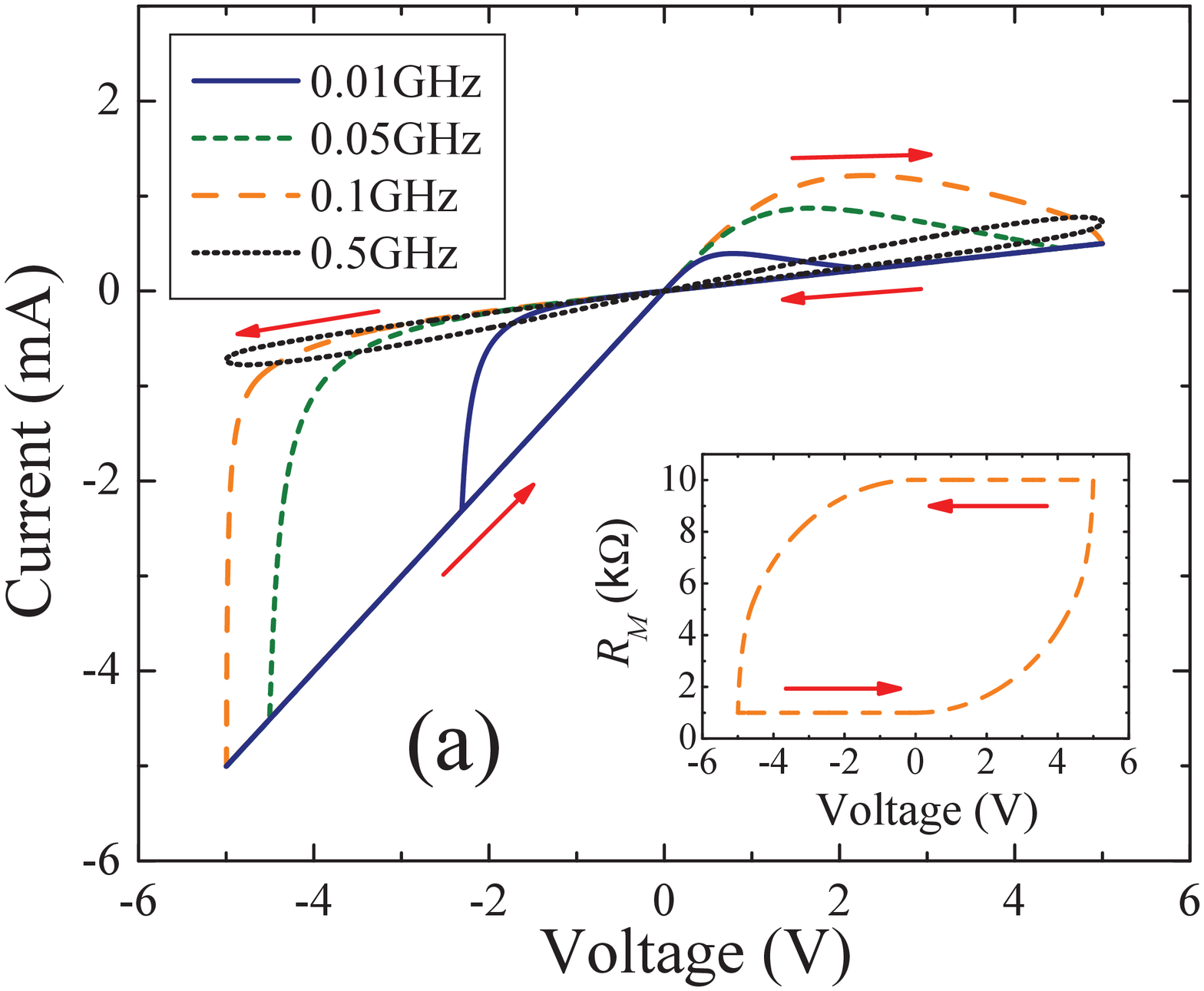}}
    \mbox{\includegraphics[width=7.00cm]{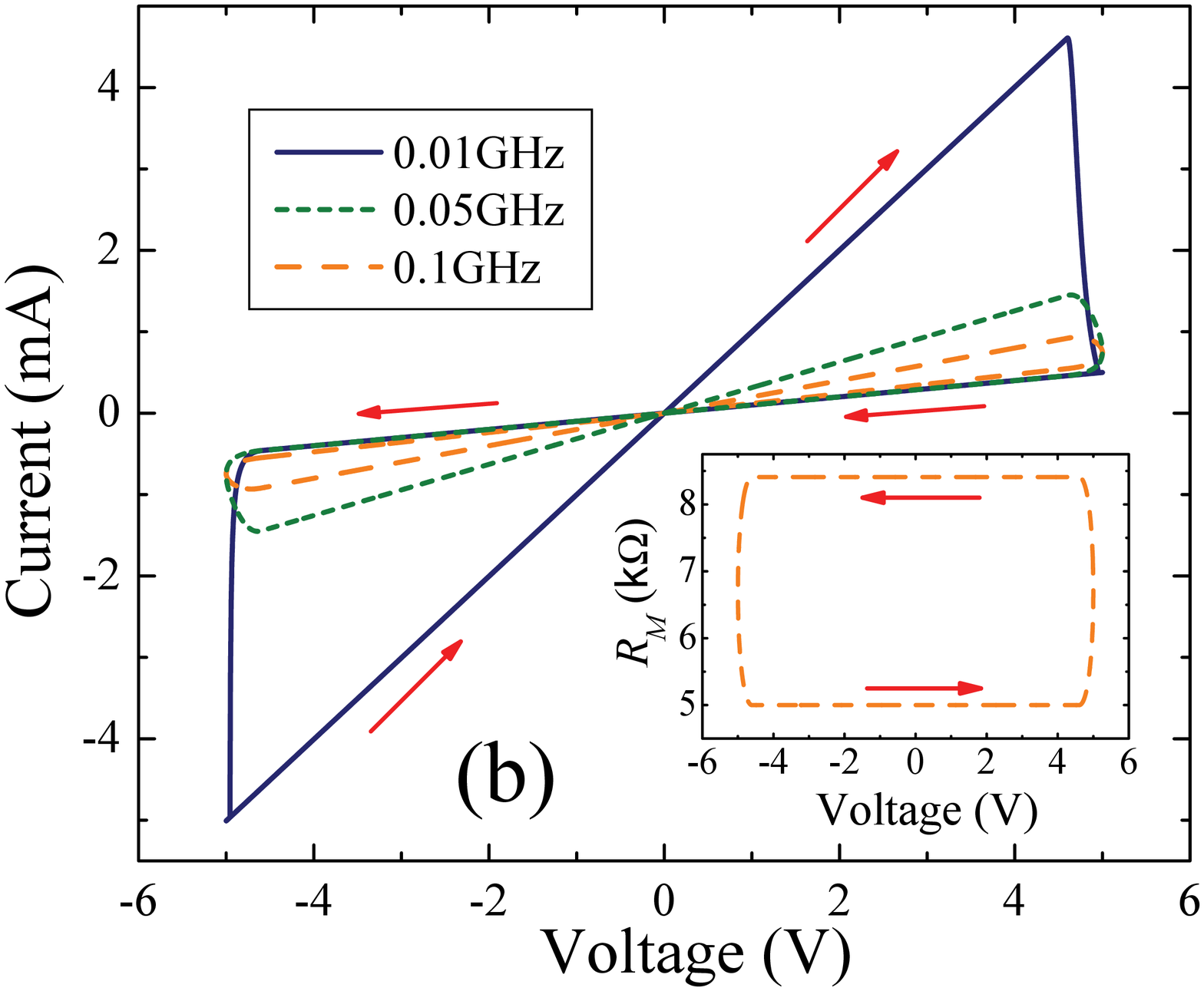}}
   }
\fcaption{\label{fig5} Frequency-dependent hysteresis loops for memristive devices with soft (a) and hard (b) thresholds.  The simulations parameters are as in Fig. \ref{fig4} except of $\nu$. The applied voltage frequencies $\nu$ are indicated on the plots.  The insets (calculated at 0.1GHz sine voltage frequency) show the memristance as a function of applied voltage. The inset in (b) demonstrates that in the case of the hard threshold the memristance changes only when the absolute value of the applied voltage exceeds the threshold voltage $V_t$.}
 \end{center}
\end{figure*}

The current as a function of voltage at several selected values of $\nu$ is plotted in Fig. \ref{fig5}. Clearly, these curves are typical frequency-dependent pinched hysteresis loops \cite{chua76a,diventra09a}. The character of the loops for memristive systems with hard and soft thresholds is slightly different. While for memristive systems with soft threshold the curve for the lowest frequency has the smallest loop span, the situation for the memristive system with hard threshold is opposite: the largest loop span occurs at the lowest frequency. This result, however, is not surprising if we take into account the fact that in the memristive system with soft threshold the change of $R_M$ occurs at lower voltages.
Moreover, the insets of Fig. \ref{fig5} demonstrate the memristance $R_M(t)$ as a function of $V(t)$ at a particular frequency. It is not difficult to notice that in the case of the memristive system with hard threshold (shown in the inset of Fig. \ref{fig5}(b)), $R_M$ changes only when $\left| V\right|$ exceeds $V_t=4.6$V .

\rengSection{Conclusions}

We have developed and tested a SPICE model of memristive devices with threshold voltage. In this model, the limiting conditions for the memristance are realized
using ternary functions which adhere more closely to the actual physical situation, compared with the window functions approach previously suggested \cite{joglekar09a}. The memristive device is realized as a
sub-circuit consisting of several elements. While the present model is based on a single internal state variable, $X$, it can be easily generalized
to more complex physical models involving several internal state variables. We would like to note that the NGSPICE model presented in Table \ref{tbl1} was included into the last distribution of NGSPICE \cite{ngspice}. Moreover, different convergence and simulation issues of memelements in SPICE will be considered in our future publication \cite{Biolek13a}. Finally, we note that threshold models of memcapacitive and meminductive systems can be implemented in the SPICE environment in a similar way.

\vspace{3em}
\noindent{\Large\bfseries Acknowledgements}

This work has been partially supported by NSF grants No. DMR-0802830 and ECCS-1202383, and the Center for Magnetic Recording Research at UCSD.

\vspace{0cm}
\begin{center}
\noindent{\Large\bfseries About Authors\dots}
\vspace{0cm}
\end{center}

\noindent\textbf{Yuriy V. PERSHIN} was born in Russia. He received his Ph.D. degree in
theoretical physics from the University of Konstanz, Konstanz, Germany, in 2002. His research interests span broad areas of nanotechnology, including physics
of semiconductor nanodevices, spintronics, and biophysics.

\noindent\textbf{Massimiliano Di Ventra} was born in Italy. He received his Ph.D. degree in
theoretical physics from the Ecole Polytechnique Federale de Lausanne, Switzerland, in 1997. His research interests are in the theory of electronic and transport properties of nanoscale systems, non-equilibrium statistical mechanics, DNA sequencing/polymer dynamics in nanopores, and memory effects in nanostructures for applications in unconventional computing and biophysics.

\end{multicols}
\end{document}